\documentclass{sig-alternate}
\usepackage[utf8]{inputenc}
\usepackage{graphicx}
\usepackage{url}
\usepackage{colortbl}
\usepackage{float}

\DeclareFixedFont{\afacc}{OT1}{phv}{m}{n}{12}

\title{Performance Evaluation of \ttlit{netfilter}}
\subtitle{A Study on the Performance Loss When Using \subttlit{netfilter} as a
Firewall}
\author{Raik Niemann, Udo Pfingst and Richard G{\afacc\"o}bel}

\begin{document}
\maketitle

\begin{abstract}
Since \emph{GNU/Linux} became a popular operating system on computer network 
routers, its packet routing mechanisms attracted more interest. This does not 
only concern ``big'' \emph{Linux} servers acting as a router but more and more 
small and medium network access devices, such as DSL or cable access devices.

Although there are a lot of documents dealing with high performance routing 
with \emph{GNU/Linux}, only a few offer ex\-pe\-ri\-men\-tal results to prove 
the given advices. This study evaluates the throughput performance of 
\emph{Linux'} routing subsystem \emph{netfilter} under various conditions like 
different data transport protocols in combination with different IP address 
families and transmission strategies. Those conditions were evaluated with two 
different types of \emph{netfilter} rules for a high number in the rule tables. 
In addition to this, our ex\-pe\-ri\-ments allowed us to evaluate two prominent 
client connection handling techniques (threads and the \texttt{epoll()} 
facility).

The evaluation of the 1.260 different combinations of our test parameters shows 
a nearly linear but small throughput loss with the number of rules which is 
independant from the transport protocol and framesize. However, this evaluation 
identifies another issue concerning the throughput loss when it comes to the 
address family, i.e. IPv4 and IPv6.
\end{abstract}

\category{C.2.6}{Internetworking}{Routers}
\category{C.4}{Performance of systems}{Measurement techniques}
\keywords{Linux, netfilter, performance}

\section{Introduction}
\noindent One of the benefits of the \emph{Linux} kernel is the availability for
nearly every technical architecture. The combination with the \emph{GNU}
operating system (often referred as \emph{GNU/Linux} or simply \emph{Linux})
makes it a good choice for router in computer networks because its memory
footprint is quite small based on the modularity of the kernel modules.

In addition to this, the \emph{Linux} kernel has out-of-the-box routing 
capabilities as well as advanced packet filter and transformation mechanisms 
which can be found in the \emph{netfilter} framework inside of the \emph{Linux} 
kernel. Quality of service based classification and priorization are available, 
too.

Along with other key features such as the big va\-rie\-ty of server software,
\emph{GNU/Linux} is now one of the most preferred operating systems especially 
for small rou\-ting devices, for example DSL or cable access devices in 
end-user environments. Another famous example for \emph{GNU/Linux} is the usage 
in wireless access routers known as \emph{DD-WRT}.

Those devices as well as ``big'' \emph{GNU/Linux} routers, for example PCs 
or servers, share the same disadvantage: the routing and filtering is based on 
software whose execution time is influenced by many factors, for example CPU, 
main memory and hardware drivers. In the worst case, the technical components 
of the router are not performant enough to process the data packets and they 
are delayed or discarded.

The main objective of this study is to evaluate the impact of \emph{netfilter} 
rules on the throughput rate per client in a distributed client-server 
application, i.e. a performance test. Although we are aware of the fact, that 
\emph{netfilter} features a variety of filter rules, we focus for our 
experiments only on the most interesting rules for router operators: rules for 
both permitting clients to pass the router (ACL) and measuring their traffic 
vo\-lu\-me (known as \emph{IP accoun\-ting}) as well as rules for regulating 
the available network bandwidth among those clients (known as QoS).

In section \ref{sec::apparatus} we describe the reasons why we did not use the 
test apparatus for this kind of performance test that is suggested in RFC 3511. 
Additionally, we spe\-ci\-fy our test apparatus and the extended set of 
possible influence parameters.

Since we used our own test apparatus, we were able to clea\-ri\-fy another 
aspect in client-server applications: the handling of client connections in the 
server component. We evaluated two widely used kinds in terms of the throughput 
rate. The first is to handle each client connection in its own thread 
(threading) and the other is the \texttt{epoll()} facility offered by the 
\emph{Linux} kernel that proclaims to be more performant and easier to 
implement. Our results along with other observations are discussed in section 
\ref{sec::results}.

\section{Test apparatus}
\label{sec::apparatus}
\noindent The test apparatus follows the guidelines described in RFC 3511 
\cite{/RFC3511/}. Basically it is a client-server architecture where a 
central gateway filters and transforms the data transmissions between the 
clients and the server.

Contrary to RFC3511, we did not use the suggested HTTP benchmark because we 
were fundamentally interested in the evaluation of a bigger number of influence 
parameters than only HTTP transactions per se\-cond. All test parameters that 
we were interested in are listed in table \ref{tab::parameters}. We developed a 
distributed application\footnote{(Link removed according to double-blind 
review process).} instead that has the same semantics like other popular command 
line benchmark tools like \emph{iperf} or \emph{netperf}, but incorperates a 
third component \emph{gateway} in the client-server concept.
\begin{figure}[!t]
\includegraphics[width=\columnwidth]{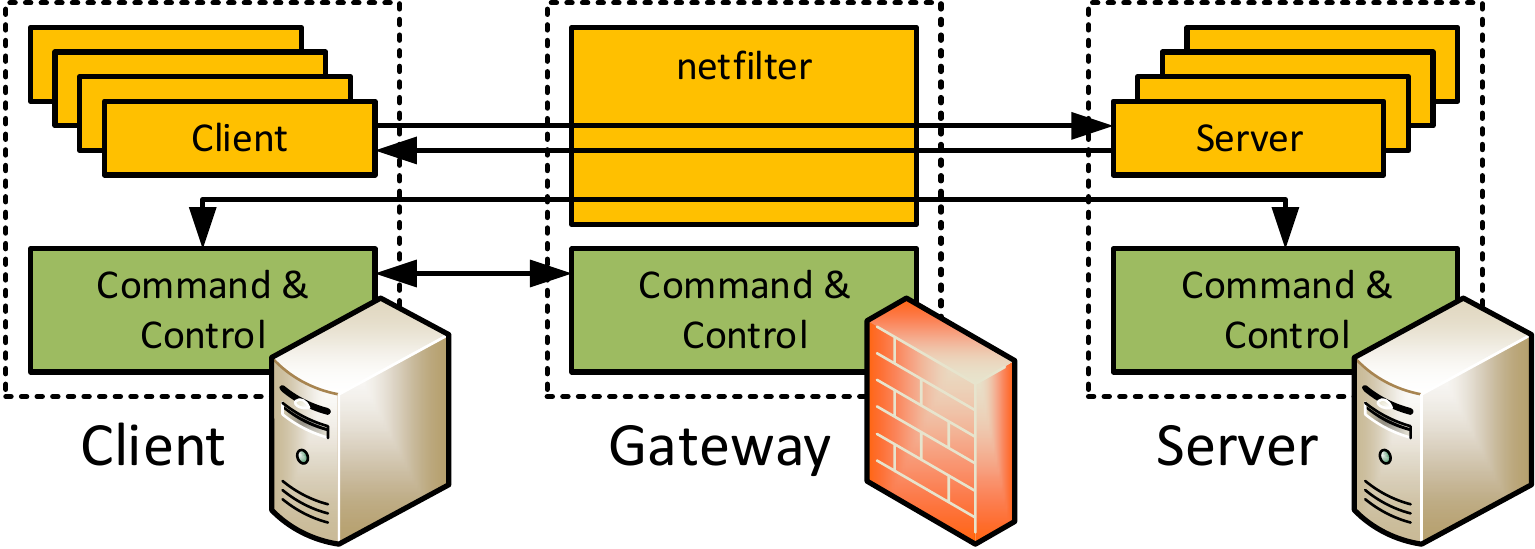}
\caption{\emph{netfilter} performance testing architecture}
\label{fig::architecture}
\end{figure}
\begin{table}[htb]
\caption{Parameters of a test case}\label{tab::parameters}
\footnotesize
\begin{tabular}{|c|p{.7\columnwidth}|}
\hline \textbf{Parameter} & \textbf{Description and tested values} \\
\hline $n$ & Number of client threads (5,10,20,40,80, 160,320) \\
\hline $t$ & Duration of the experiment (100s) \\
\hline $f$ & Frame size (either fixed [64,128,256,512, 1024] or ranged between 
64 and 1024) \\
\hline $P$ & Transport protocol for the transmission (either TCP, UDP or SCTP) 
\\
\hline $A$ & Address family (either IPv4 or IPv6) \\
\hline $T$ & Server component uses threads for handling the client connections 
(only valid for stream oriented protocols) \\
\hline $F$ & \emph{netfilter} rule generation per client thread: 0 for plain 
forwarding, 2 for up- and download and 4 for additional QoS marks \\
\hline
\end{tabular}
\end{table}

\noindent As shown in figure \ref{fig::architecture}, our application consists 
of three independant command line tools. The communication between those three 
components can be divided into two parts: a) the control connection between the 
client and the gateway as well as the server component and b) the data 
connections between the client threads and the server thread(s). The control 
connection is used for sending the test parameters to the components and 
(once they configured themselfes according to the parameters) to signal the 
start and the end of the specific experiment.

\begin{figure}[!t]
\centering
\includegraphics[width=\columnwidth]{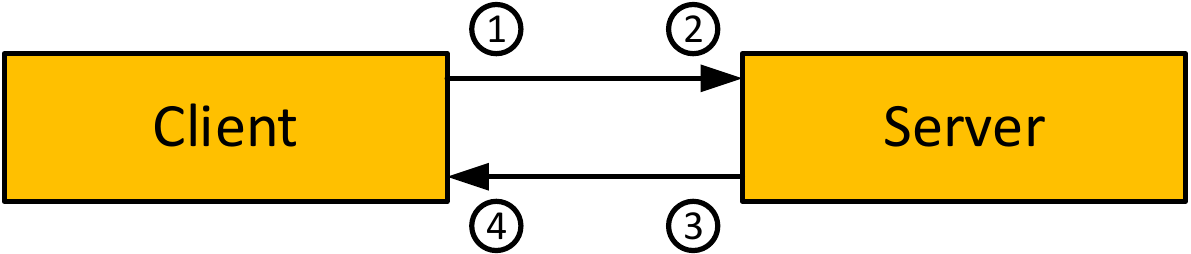}
\caption{Measurement points of the sending/receiving cycle}
\label{fig:measurepoints}
\end{figure}

\noindent Every experiment follows the same steps:
\begin{enumerate}
\item The test parameters are given as command line pa\-ra\-me\-ters when the 
client component starts. The parameters are validated and transferred to the 
gateway component. The gateway component configures the \emph{netfilter} 
subsystem according to the submitted test parameters by inserting appropiate 
filter rules as specified in test parameter $F$.
\item When the gateway components signals its readiness for the configuration, 
the client component submits the test configuration to the server component. 
The server component then awaits any client connections by opening a server 
socket. When this socket is successfully opened and bound, the server component
signals its readiness back to the client component.
\item The client component initializes and executes $n$ client threads. They
subsequently connect to the server component. According to the parameter $T$ 
the server component handles each of the client connection in a) its own thread 
or b) in a single thread using the \texttt{epoll()} facility.
\item Once all client threads are connected, they begin to send and to receive 
data packets according to the test parameters $P$,$A$ and $f$. Every of the $n$ 
client-server-connections has its own independant sending/receiving cycle as 
depicted in fi\-gu\-re \ref{fig:measurepoints}: the client sends a specific 
amount of data (measurement point 1), the server thread receives the data 
(measurement point 2) and echos it back to the client thread (measurement point 
3). The client thread finally receives the data (measurement point 4).
\item When the test duration $t$ is reached, the client threads get a signal to 
end the current sending/re\-cei\-ving cycle, to disconnect from the server 
component and finally to end. The client component instructs the server and 
then the gateway component to restore the system state that existed before the 
ex\-pe\-ri\-ment.
\end{enumerate}
For each of the four measurement points as shown in fi\-gu\-re
\ref{fig:measurepoints}, several values were recorded. For measurement point 1
and 3 the number of successful sent/unsent data frames and the frame sizes were
saved. The term ``unsent'' in this context means that a data frame could not be
send successfully within a specific timeout (500 ms).

For measurement point 2 and 4 the number of received data frames as well as the 
frame size and the result of the validation were saved. Please note that a read 
timeout was possible, but not used. For the validation process every data frame 
sent by a client was filled with a data record that contains the following 
information: a) the number of the client in the range from 1 to $n$ b) the 
chosen frame size according to test parameter $f$ and consecutive sequence 
number starting with 1 and raised with every send/receive cycle. This allows to 
validate if a received data frame belongs to the associated sender and the data 
was successfully transmitted by comparing the received amount of data with test 
parameter $f$. In addition to this, it allows to detect ``gaps'' in the 
sending/receiving cycle.

All those recorded values formed the basis for our evaluation.

\subsection{Test series}
\noindent We composed three test series based on the test pa\-ra\-me\-ter $F$ 
(refer to table \ref{tab::parameters}):
\begin{enumerate}
\item \emph{Plain forwarding}: this test series only makes use of 
\emph{netfilter}'s forwarding capabilities. This means that the gateway 
component is instructed to forward all data transmission between the client and 
the server component without any limitations, i.e. no \emph{netfilter} rules 
were inserted.
\item \emph{Simple up- and download rules}: this test series is like the first 
one but the gateway component inserts a upload and a download \emph{netfilter} 
rule per client thread. The rules simply checks the IP addresses and the 
protocol to test\footnote{To be more precise: ``\texttt{iptables -A 
FORWARD -s <client> -d <server> -p <protocol> -j ACCEPT}'' for the upload and 
vice versa for the download direction.}. In total $2 \cdot n$ rules are active 
for a specific experiment. At the end \emph{netfilter} is instructed to discard 
any other data packet that does not conform with the inserted rules. This is 
done by setting the policy of the specific rule table to drop anything that was 
not matched by any existing rule.
\newpage\item \emph{Simple up- and download rules as well as QoS marks}: this 
test series does the same as the se\-cond one but additionally inserts 
\emph{netfilter} rules per client thread that are responsible to tag in- and 
outgoing network data packets with a QoS mark\footnote{The rule template for 
this is ``\texttt{iptables -t mangle -A PREROUTING -s <client> -d <server> -p 
<protocol> -j MARK --set-mark <QoSmark>}'' for the upload and vice versa for 
the download direction.}. Those marks can be used within the \emph{iproute2}
uti\-li\-ty collection to manipulate the QoS subsystem of the \emph{Linux} 
kernel. In total $4 \cdot n$ \emph{netfilter} rules are inserted for a specific 
experiment. Please note that the QoS subsystems of all three test machines were 
not modified and used the default (\emph{pfifo\_fast}, a simple packet 
first-in-first-out queue with almost no overhead).
\end{enumerate}
The results of the first test series served us as a baseline for the other two. 
During the experiments the hardware metrics were recorded, e.g. CPU and main 
memory usa\-ge.

\subsection{Test machines characteristics}
\noindent The machine for the client component has two \emph{AMD Opteron 870} 
CPUs with 4 cores each and 2 GHz frequency. The machines for the gateway and 
server component have two \emph{AMD Opteron 890} CPUs with 4 cores each and 2.8 
GHz frequency. Each of the three machines have 32 GByte of main memory (DDR2, 
ECC error correction).

All test machines used a recent \emph{GNU/Linux} distribution (Ubuntu 14.04 in 
the 64 bit server edition) as an operating system with a recent \emph{Linux} 
kernel (3.13-03). All unnecessary services were turned off.

\subsection{Network configuration}
\noindent Each of the three machines used for the tests has a 4-port network 
adapter with two \emph{Intel 82546EB} chipsets. This allows four physical GBit 
connections. The gateway machine is dual-homed with a physical GBit connection 
to each the client and server component machine. Each the client and server 
component has its own IP network.

Since the gateway machine is dual-homed, it can connect both networks and uses 
\emph{netfilter} to route, to filter and to transform the data transmissions 
between the two networks.

All settings that were available for the tested transport protocols and 
address families were left to their defaults. Although they offer the potential 
to raise the processing performance, the complexity in conjunction with our 
test parameters was too high.

\newpage\section{Discussion of the test results}
\label{sec::results}
\noindent We executed every test series three times and all shown results use 
the mean value; the variance was uniformly low. In total we executed 3.780 
single experiments.

\subsection{General observations}
\begin{figure}[!t]
\includegraphics[width=\columnwidth]{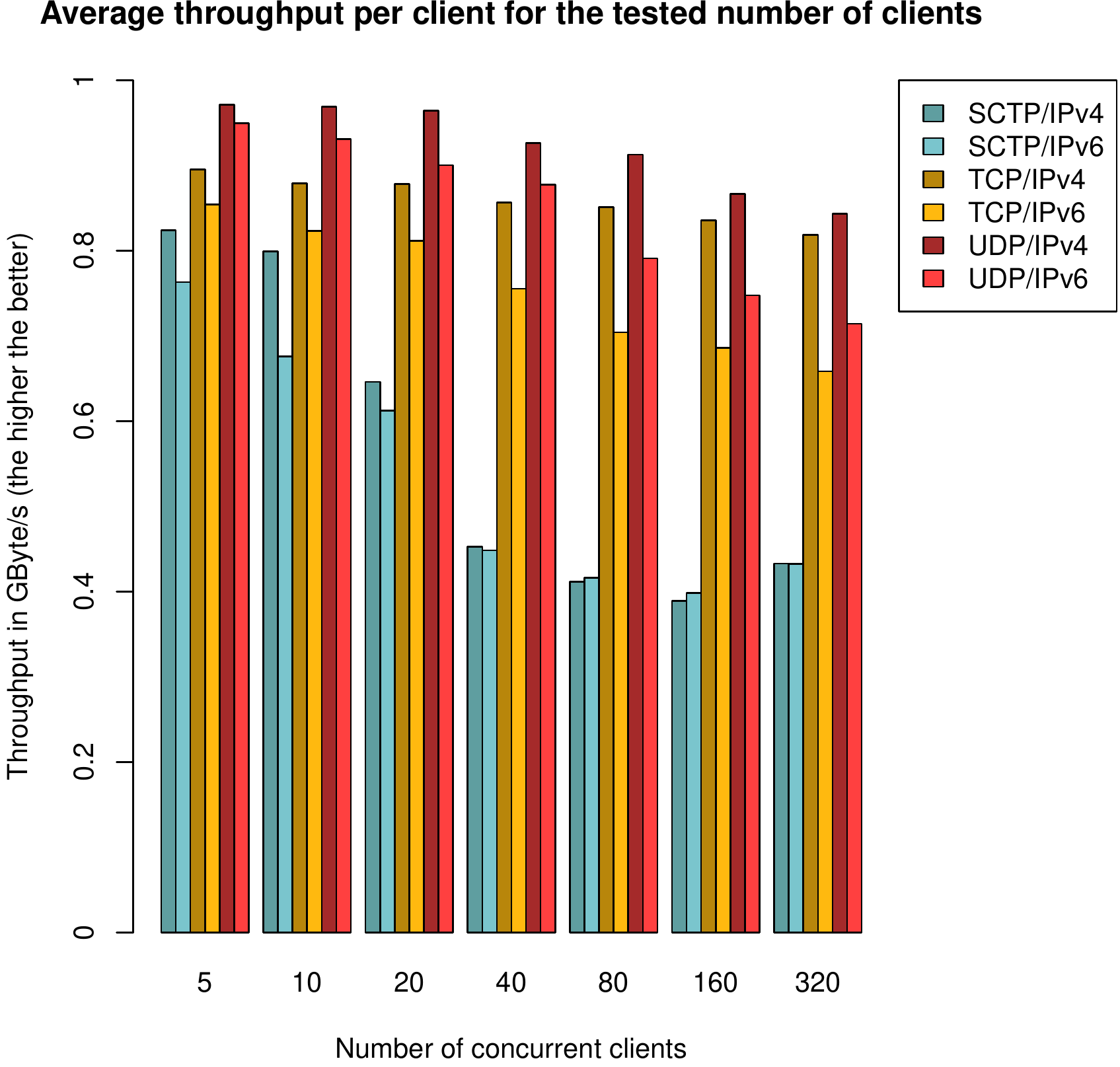}
\caption{Average throughput rate per client for the tested number of clients}
\label{fig::throughput-client}
\end{figure}

\begin{figure}[!t]
\includegraphics[width=\columnwidth]{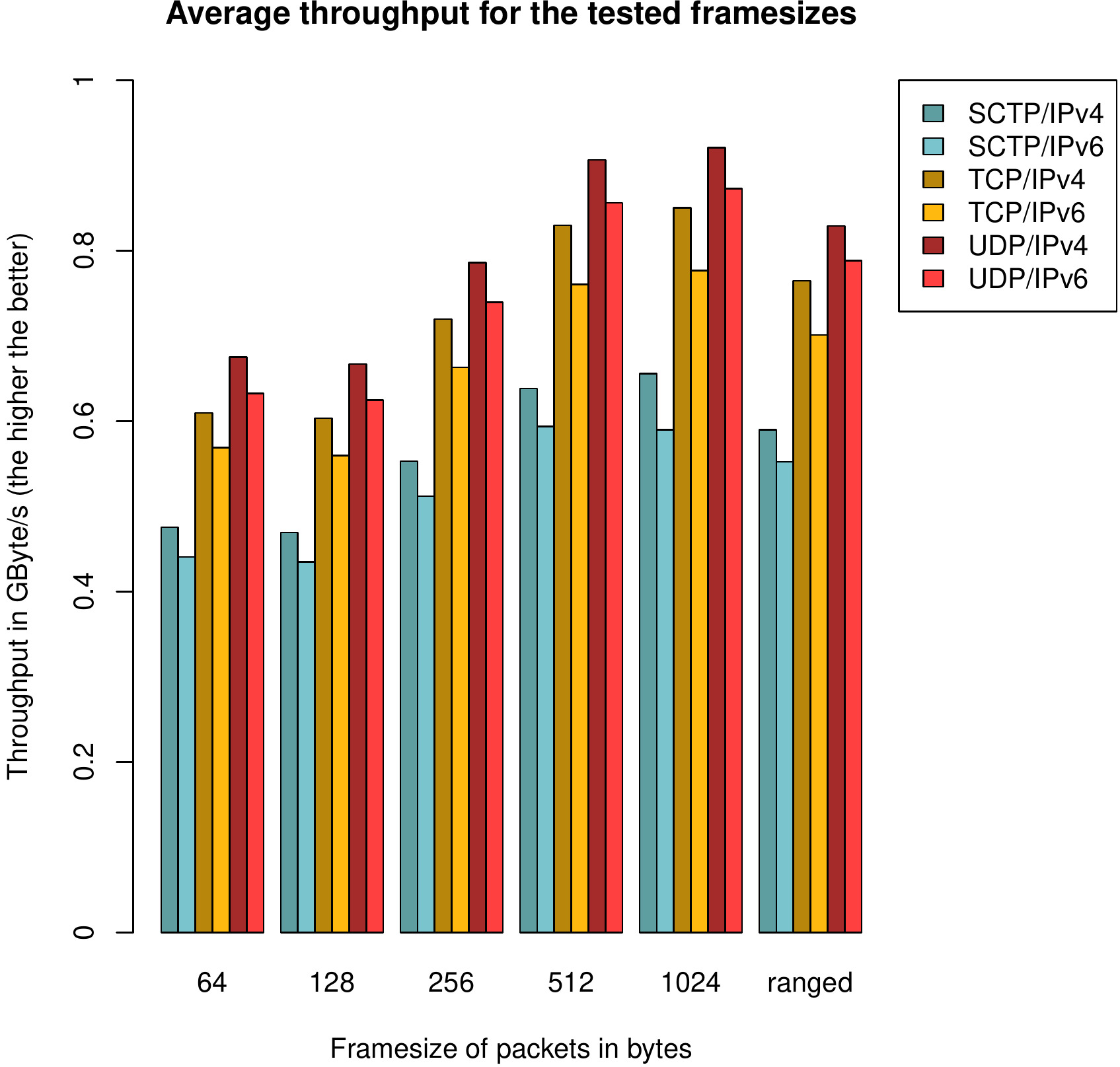}
\caption{Average throughput rate per client for tested frame sizes}
\label{fig::throughput-framesize}
\end{figure}

\noindent All three test series gave us a first impression of the throughput 
rate for the tested protocols and address families. The average throughput rate 
for all 3.780 experiments is depicted in figure \ref{fig::throughput-client} 
and \ref{fig::throughput-framesize}. Both show the results for the tested 
address families and scaled to the potential transmission maximum of 1 GBit per 
second. 

The first figure shows the throughput rate grouped by the tested number of 
concurrent client threads. This way it is possible to estimate the average 
throughput for any application where the number of clients are known. Please 
note that the shown throughput rates already include the decrease resulted by 
\emph{netfilter}'s filtering and routing. As visible in figure 
\ref{fig::throughput-client}, the average throughput rate is quite stable but 
decreases with a higher number of concurrent clients. The only exception is 
SCTP 
where the throughput rate is surprisingly higher for 320 concurrent clients 
than 
for 80 and 160.

The latter figure \ref{fig::throughput-framesize} shows the throughput rate
grouped by the tested frame sizes. This figure also include all experiments 
where \emph{netfilter} rules were involved. Unsurprisingly the throughput rate 
increases with a higher frame size. The general case is shown in the last bar 
group labeled ``ranged''. In this case the frame size was randomly 
chosen\footnote{A \emph{Mersenne} random number generator was used.} in a range 
between 64 and 1024 before every send/receive cycle in every client thread.

We can confirm the widely known fact that SCTP in terms of throughput is slower 
than TCP which is slower than UDP. Our results show that SCTP is in average 
32.65 percent slower than TCP (minimum/maximum difference: 9.28 and 48.23 
percent) for all experiments. TCP however is in average 8.42 percent slower 
than UDP (minimum/maximum difference: 5.18 and 10.64 percent). 

Our test results also showed that the throughput rate for IPv6 is noticeable 
lower than for IPv4. All tested protocols using IPv4 are in average 9.22 
percent faster than with IPv6 (minimum/maximum difference: 4.59 and 13.8 
percent).

As mentioned before, we recorded the available hardware usage statistics during 
all experiments. Compared to the statistics for our router machine in its idle 
state, the impact of the \emph{netfilter} routing during the experiments in 
average is marginal. In fact, this depends on the utilized network adapters and 
the system drivers. Our network adapters featured a special network processor 
that massively reduced the CPU load by va\-li\-da\-ting incoming network data 
packets natively, e.g. calculating checksums and verifying packet headers, 
which 
is otherwise done by the operating system.

\subsection{Impact of netfilter}
\noindent As stated in the previous section, the first test series (without any 
\emph{netfilter} rules) served us as a baseline for the other two that we 
executed (with different numbers of \emph{netfilter} rules).

\begin{figure}[!t]
\includegraphics[width=\columnwidth]{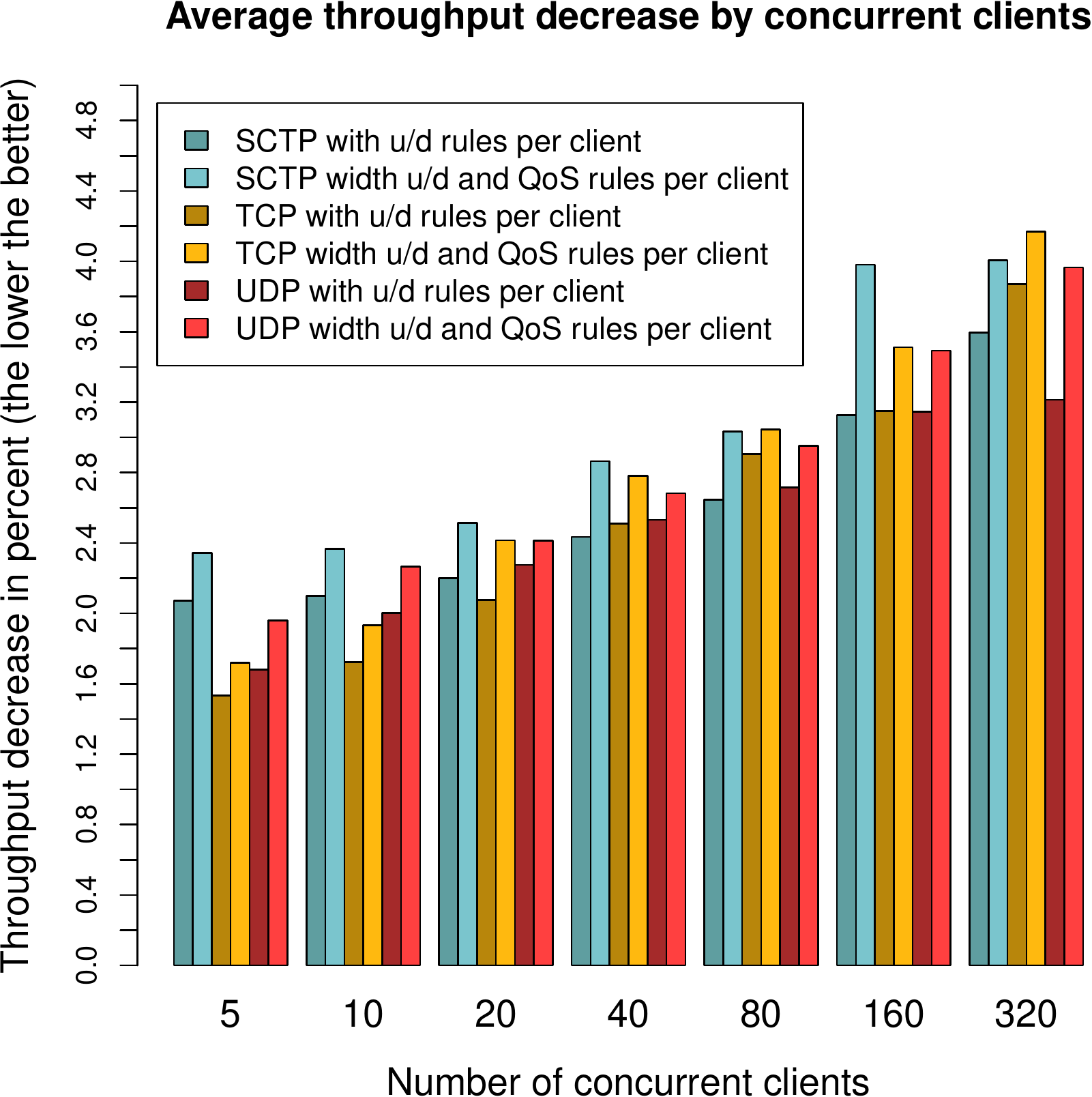}
\caption{Average throughput decrease per client for IPv4 when \emph{netfilter} 
rules are active (u/d = up- and download)}
\label{fig::diff4}
\end{figure}

\begin{figure}[!t]
\includegraphics[width=\columnwidth]{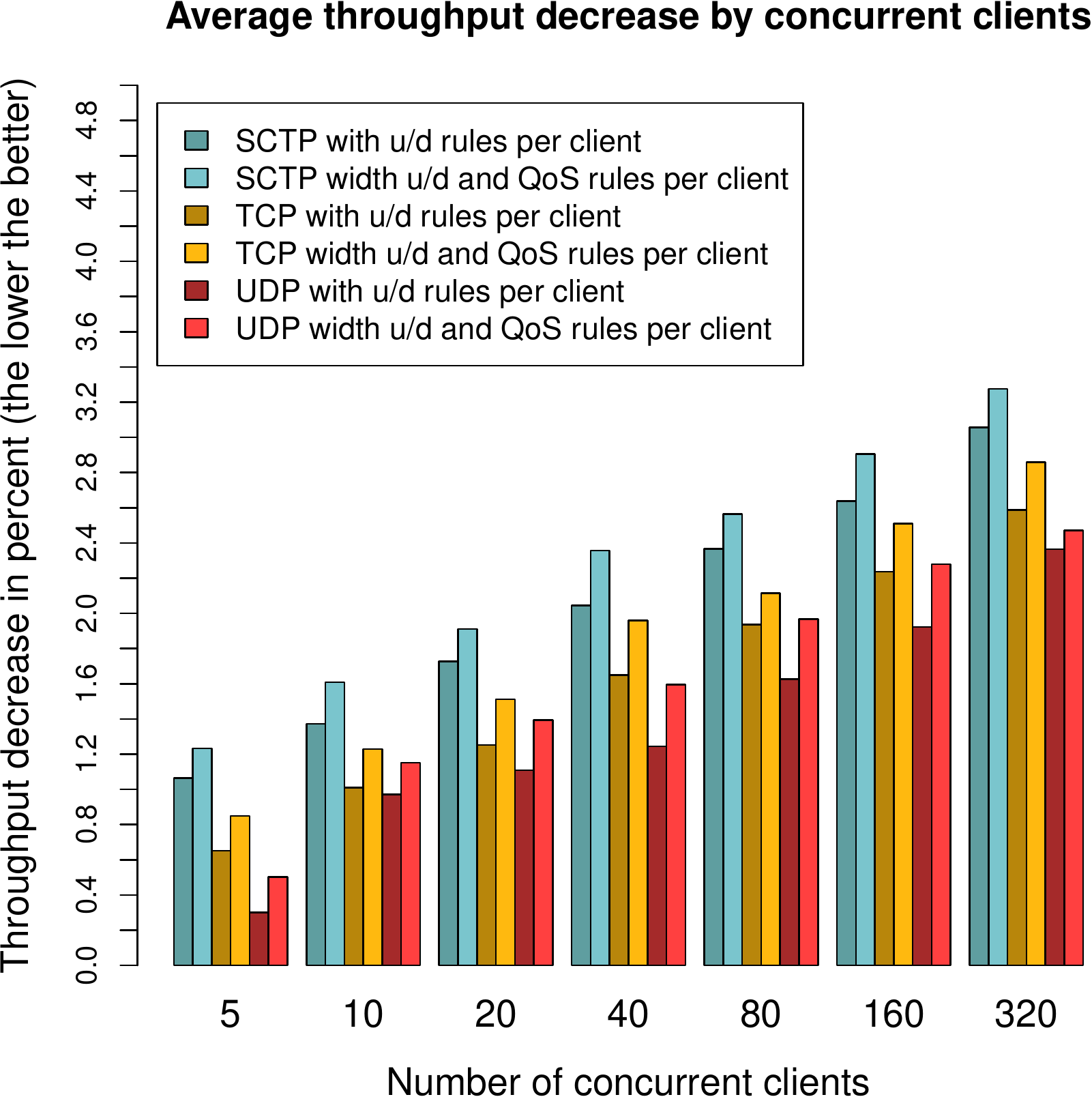}
\caption{Average throughput decrease per client for IPv6 when \emph{netfilter} 
rules are active (u/d = up- and download)}
\label{fig::diff6}
\end{figure}

\newpage
\noindent For the second and third test series, we calculated the difference 
with the first one. The results showed a decrease of 2.25 percent in average for 
all experiments where \emph{netfilter} rules were involved. We summarized the
average throughput decrease in figure \ref{fig::diff4} (IPv4) and 
\ref{fig::diff6} (IPv6). These figures show the average throughput decrease 
grouped by the tested client thread numbers and additionally for every tested 
protocol and number of active \emph{netfilter} rules per client.

As depicted in figure \ref{fig::diff4} and \ref{fig::diff6}, the decrease is 
different for the tested address families: the decrease for IPv6 is lower than 
for IPv4 (2.71 vs. 1.79 percent in average). By considering the decrease 
percentages as a function of the number of inserted \emph{netfilter} rules, we 
calculated the gradient for each tested protocol and address family. In average 
the gradients are \emph{nearly constant}. This can be barely seen on figure 
\ref{fig::diff4} and \ref{fig::diff6} because the x-axis is not linearly 
scaled. To confirm the nearly constant impact of \emph{netfilter} on the 
throughput rate, we reviewed our test results with respect to the number of 
routed data packets between the client and server thread(s) rather than the 
throughput rate. The review also proves our main findings:
\begin{enumerate}
\item \emph{netfilter}'s performance in terms of throughput is independant from 
the used transport protocol, fra\-me size and address family as long as simple 
\emph{netfilter} rules are active
\item the throughput loss increases roughly linear with the number of inserted 
(simple) \emph{netfilter} rules although this loss is quite insignificant
\end{enumerate}
The second main finding shown above allowed us to express the throughput loss 
per \emph{netfilter} rule: one can assume a throughput loss of 0.05 percent for 
any (simple) IPv4 rule and 0.03 percent for any (simple) IPv6 rule.

\subsection{Client handling techniques}
\noindent The last objective of this study was to evaluate the client 
handling techniques in a client-server application: the server component was 
instructed to handle the data transmissions of stream-oriented clients either in 
a separate thread per client or in a single thread using \texttt{epoll()}.

\begin{figure}[!t]
\includegraphics[width=\columnwidth]{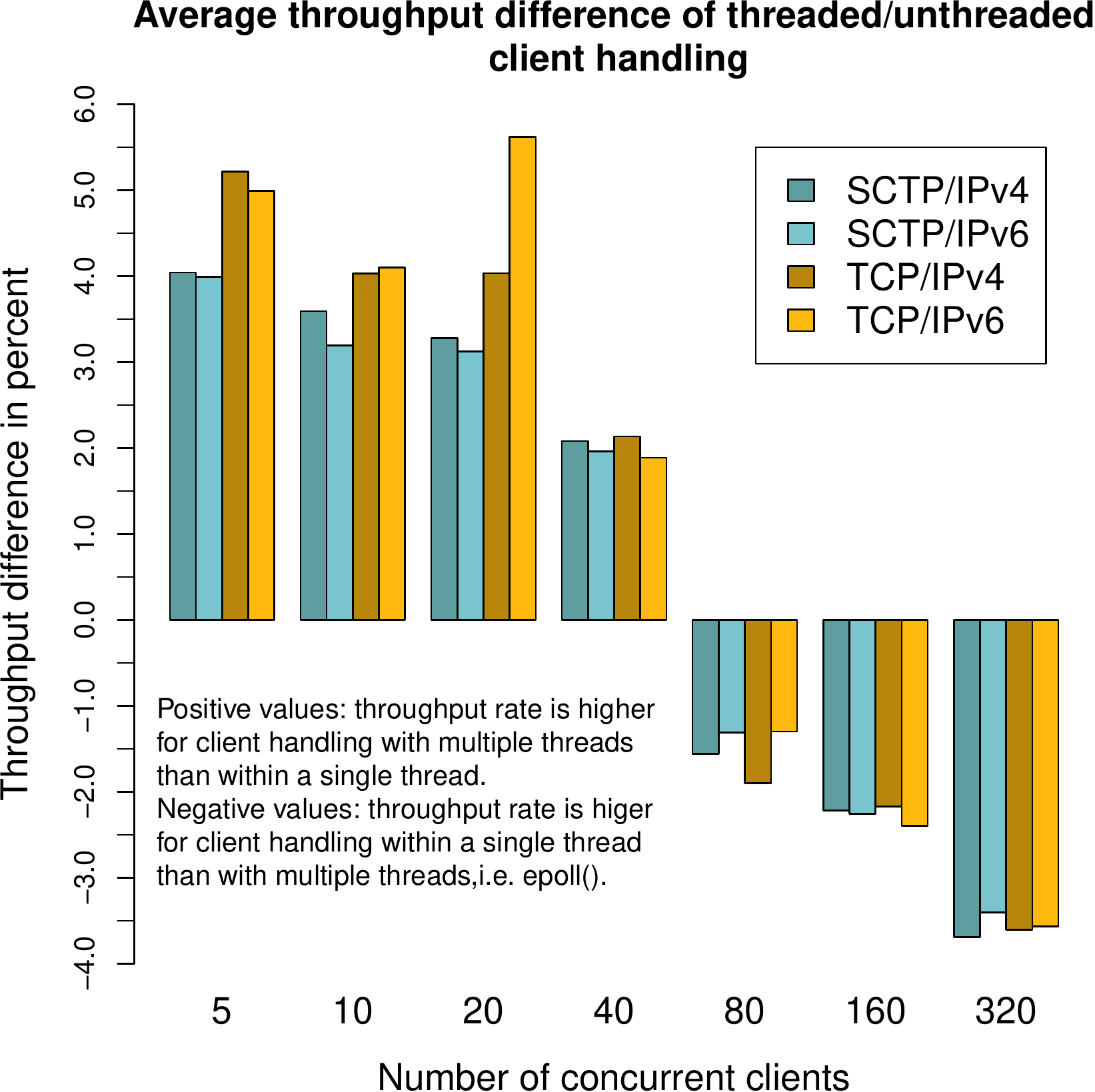}
\caption{Average throughput difference between client handling with multiple 
threads and within a single thread}
\label{fig::threading}
\end{figure}

To consider the differences in the throughput between those two client handling 
techniques, we only used the experiment results of the first test series and 
only for SCTP and TCP as well as for both address families. The average 
difference is illustrated in figure \ref{fig::threading} as a percentage 
between the threaded and unthreaded technique.

This figure clearly indicates that there is a turning point which technique 
offers a higher throughput rate for a specific number of client connections to 
handle in a server process.

\noindent In our experiments this turning point was around 40 concurrent client 
connections. The technical specifications of our test machine executing the 
server component states the native handling of 32 concurrent threads. This 
brought us to examine the system usage statistics that were recorded during the
experiments. We noticed a significant increase of the number of context 
switches for our experiments with more than 40 concurrent threads. A context 
switch takes place when the operating system saves the current state of a 
process or thread for a later execution in favor of the execution of another 
process or thread. This storing/restoring of contexts is quite expensive in 
terms of computation time and can cause the system to slown down. In 
contrast the same experiments with 40 or more client connections that were 
handled via \texttt{epoll()} in a single thread did not show this impact.

In summary we recommend to use the \texttt{epoll()} facility of the \emph{Linux}
kernel in a client-server architecture in general. The reason is the better 
scalability compared to a client hand\-ling with threads for a higher number of 
client connections. Although the throughput rate is higher when threads are 
used, the rate difference is not significantly higher compared to the 
hand\-ling with \texttt{epoll()}. In addition to this, an application using 
\texttt{epoll()} can prevent the operating system from unnecessary context 
switches that also effects other concurrent applications.

\section{Related work}
\noindent In March 2000, the \emph{netfilter} routing subsystem was mer\-ged 
into the \emph{Linux} kernel as the succesor of the former subsystem 
\emph{ipchains}. The first performance evaluations regarding this new subsystem 
were made by \emph{Hartmeier et al.} and \emph{Podey et al.} 
(\cite{hartmeier2002design}, \cite{podey2003network}). The comparision 
bet\-ween their results concerning the throughput rate with ours for a high 
number of \emph{netfilter} rules indicates the same correlations but also 
illustrates the improvements in the \emph{Linux} kernel and \emph{netfilter} 
subsystem since then.

Further publications dealt with the architecture of \emph{netfilter} to raise 
the performance. The \emph{netfilter} rules are organized in tables that are 
consulted according to the state of a network data packet. In general, rule 
evalation is done sequentially in each table. \emph{Lyu et. all} as well as 
\emph{Fulp} (\cite{lyu2000firewall}, \cite{fulp2005optimization}) classified 
rules for a later elimination of unnecessary rules. This decreased the overall 
effort to inspect a data packet within \emph{netfilter} and lead to a better 
throughput.

In addition to this, user-defined sub-tables can be crea\-ted in each of 
\emph{netfilter}'s pre-defined tables and can be used as a target for a rule. 
This allows the segmentation of the rule evalation. \emph{Fulp et all.} 
\cite{fulp2005optimization} showed that the rules can be organized as a trie to 
achive a faster rule evalation.

In \cite{acharya2006simulation}, \emph{Acharya et. all} collected real-world 
firewall rule sets of tier-1 internet service providers and the associated 
usage statistics to form a model for analyzation. This model was later used to 
improve the rule sets in order to increase the throughput.

\emph{Accardi et. all} \cite{accardi2005network} used a special expansion card 
with a programmable network processor to relocate the network data packet 
inspection in combination with a \emph{netfilter} module for this purpose. 
Their 
results show a tremendous increase of packet processing in a worst-case 
scenario, e.g. a Denial-of-service attack. In this attack scenario, the 
\emph{netfilter} router faces a massive amount of invalid packets. 
\emph{Accardi 
et. all} demonstrated that their setup of programmable network processor and 
corresponding \emph{netfilter} module can prevent the effects of a 
Denial-of-service attack.

\section{Conclusion and future work}
\noindent In this study we presented the results of our experiments studying 
the impact of \emph{netfilter} on the throughput rate. We tested different 
combinations of transport protocols, address families and frame sizes for an 
increasing number of \emph{netfilter} rules. In summary we found out that the 
throughput loss does not depend on those pa\-ra\-me\-ters. The throughput loss 
is also quite insignificant and rises roughly linear with the number of rules. 
Our experiments showed an average throughput loss of 0.05 percent for any 
(simple) IPv4 rule and 0.03 percent for any (simple) IPv6 rule.

In addition to this, we evaluated two prominent client handling strategies for 
the server component in a client-server application. We proved that up to a 
certain point a client handling with threads offer a higher but only slight 
performance gain compared to the counterpart using the \texttt{epoll()}
facility. After this point the thread management is too expensive in terms of 
computation time and causes the throughput rate per thread to degrade. The 
\texttt{epoll()} facility in contrast does not show this behaviour.

With the introduction of \emph{nftables} as the designated successor 
of the current \emph{iptables}, a performance gain is expected (although it is 
based on \emph{netfilter}, too). As soon as \emph{nftables} becomes stable, we 
will redo our experiments using this tool.
\bibliographystyle{abbrv}
\bibliography{linux-nat}

\end{document}